\documentclass[twocolumn,showpacs,preprintnumbers,amsmath,amssymb,pra]{revtex4}
\usepackage{graphicx}
\usepackage{dcolumn}

\begin{document}

\title{Role of seeding the cavity of a two-photon correlated emission laser with thermal light}

\author{Sintayehu Tesfa}
\affiliation{Max Planck Institute for the Physics of Complex Systems, N$\ddot{o}$thnitzer Str. 38, 01187 Dresden, Germany\\Physics Department, Dilla University, P. O. Box 419, Dilla, Ethiopia}

\date{\today}

\begin{abstract} A study of the evolution of two-mode squeezing, entanglement and intensity of the cavity radiation of a two-photon correlated emission laser initially seeded with a thermal light is presented. The dependence of the degree of two-mode squeezing and entanglement on the intensity of the thermal light and time is found to have more or less a similar nature, although the actual values differ specially in the early stages of the process and when the atoms are initially prepared in nearly 50:50 probability to be in the upper and lower energy levels. Particularly, seeding the cavity turns out to spoil the nonclassical features significantly in the vicinity of $t=0$.  It is also shown that the mean photon number in a wider time span has a dip when mode $b$ is seeded, but a peak when mode $a$ is seeded. Moreover, this study asserts that the effect of the seeded light on the nonclassical features and intensity of the cavity radiation is eroded with time by the pertinent emission-absorption mechanism which can be taken as an encouraging sign in practical utilization of this quantum system as a source of bright entangled light.\end{abstract}

\maketitle

 \section{Introduction}

A three-level cascade laser  has received a considerable attention over the years in connection with the strong correlation between the modes of the generated radiation that leads to a substantial degree of nonclassical features \cite{pra74043816,pra77013815,pra79013831,oc283781,prl601832,pra415179,pra484686,prl94023601,pra79063815,pra77062308,jpb41215502}. One of the various possible schemes, in this regard,  is a correlated emission laser  in which the atomic coherence is induced by initially preparing the atoms in the coherent superposition of the atomic energy levels between which a direct spontaneous transition is electric dipole forbidden (description of the model can be found for instance in \cite{pra74043816, pra77013815,jpb402373}). This quantum system has been thoroughly studied elsewhere for different situations  when the cavity is assumed to be initially in a vacuum state \cite{pra74043816,pra77013815,jpb402373,jpb42215506}. 

It is observed in these works that a strong robust entangled light can be produced by manipulating the rate at which the atoms are injected into the cavity and the initially prepared coherence.  In addition, the mean photon number  was found to be higher where the degree of entanglement is larger \cite{jpb42215506}. However, the assumption pertaining to no radiation initially in the cavity appears to pose a serious challenge when weighed solely from technical point of view since it is practically impossible to build a cavity entirely in the vacuum state. Even if one is determined to do so, it requires a significant effort and resource to cool it at that level. Undoubtedly, this recurs additional cost leaving the emerging technical difficulty aside. 

To circumvent the prevailing predicament somehow, recently there are efforts in assuming the cavity to contain a coherent light with certain number of photons when the atoms  are taken to be pumped externally with another coherent light \cite{pra79013831,pra75062305}. It was predicted that the coherent light increases the mean photon number and enhances the dependence of the entanglement on the phase fluctuations. At this juncture, there is one outstanding issue to be addressed in relation to having three coherent lights (coherent light with different frequency) in the cavity at the same time. 

From naive understanding of atom-radiation interaction, it would be reasonable to expect the atoms to respond to the coherent lights in the same manner. Consequently, as long as there is a sufficient amount of resonant coherent light  in the cavity, there is a good chance of initiating a two-step pumping process. This provides  another root for inducing a coherent superposition between the lower and the upper energy levels that amounts to establishing a similar effect as the external coherent radiation. In light of this, the general tendency of designating the observed result to the pumping radiation and the initial coherent radiation in the cavity seems to beg for a more serious indepth consideration than the mere inclusion of the coherent light as initial state in the characteristic function can warranty. 

One of the most realistic proposals to study the effect of the light in the cavity at the beginning on the generated radiation may be heating the cavity to certain particular temprature using ordinary light source and then inject the initially prepared atoms to it. This can be done by placing the cavity in a bath of ordinary light allowed to leak into it via the coupler mirrors and wait until thermal equilibrium is reached or put the light source in the cavity until a wall temprature of interest has reached and then turn it off prior to the injection of the atoms. The former scenario is the usual consideration of the thermal reservoir treated elsewhere \cite{jpb402373}. In this case, the thermal light is found to increase the mean photon number, but degrade the two-mode squeezing and entanglement. However, the analysis of the  latter is yet to be addressed. 

With this understanding, in this communication, the effect of the thermal light initially seeded in the cavity as in the second option on the statistical properties and quantum features of the cavity radiation is studied where the thermal light is conveniently represented by the chaotic state. Although the coupling of the quantum system to the environment invariably disparages the entanglement and other quantum features, the effect of decoherence due to thermal light is not taken into consideration. Nevertheless, the environment is assumed to be made of vacuum modes whose fluctuations would be included in the usual manner. 

Since the approach followed is presumed to be adaptable for arbitrary reservoir with a minimum effort, the thermal light outside the cavity has not been given attention to deter the confusion. Basically, the cavity constructed in an open space at room temprature and then placed in an experimental setup can be considered as the cavity that contains chaotic light. In addition to this, when it becomes practically impossible to pump out all radiation from the cavity, so that it would be in a vacuum state, the remnant light can also be treated as a chaotic light. Whatever the case may be, studying the effect of light in the cavity before the injection of atoms is envisioned to be of a paramount importance in utilizing the generated radiation for particular purposes like teleportation and quantum information processing as anticipated. 

To achieve the intended goal, the procedure by which the chaotic light in a cavity can be integrated with the three-level atomic system would be formulated. To do so, the general characteristics of thermal light is outlined in the way that it can be salvaged for present work. Then the equations of evolution for the three-level laser for an arbitrary initial radiation would be obtained where the contribution of the thermal light enters via this initial condition. In the process, it is assumed that the atoms are unable to distinguish among cavity radiations other than their frequency whereby the thermal light can also be absorbed by the atoms. In light of this, the effects of initially heating the cavity (thermal seeding) on the achievable two-mode squeezing, entanglement and mean photon number is investigated by making use of the correlation properties of the chaotic light. 

\section{Sketchy descriptions of the involved systems}

\subsection{Chaotic light}

It is a well established fact that the radiation generated by conventional sources of light can be taken as the best example of chaotic light. For instance, the light produced by a laser machine operating below threshold exhibits chaotic nature. For the sake of convenience, let us consider the thermal light in a cavity maintained at equilibrium with the walls  at temperature $T$. The density operator $\hat{\rho}$ for the thermal light can be related to the entropy by von Neumann relation \cite{von}
\begin{align}\label{s01}S=-\kappa_{B} \;Tr(\hat{\rho}\;ln\hat{\rho}),\end{align} where $\kappa_{B}$ is the Boltzmann constant, $Tr$ stands for trace operation and $\hat{\rho}$ is the density operator. 

On the basis  that at equilibrium entropy should be maximum and $Tr(\hat{\rho})=1$, the density operator for the chaotic light is related to the temprature by
\begin{align}\label{s02}\hat{\rho}={e^{-\hat{H}/\kappa_{B} T}\over Tr(e^{-\hat{H}/\kappa_{B} T})},\end{align} where $\hat{H}=\omega\hat{a}^{\dagger}\hat{a}$ is the Hamiltonian of the radiation field when the zero point energy is shifted by ${\omega\over2}$ and $\hbar =1$ for convenience. 

The same density operator can be described in the number state applying the completeness relation for number state $\hat{I}=\sum_{n=0}^{\infty}|n\rangle\langle n|$ and the orthonormality condition $\langle n|m\rangle=\delta_{nm}$ as
\begin{align}\label{s03}\hat{\rho}=\sum_{n=0}^{\infty}{\bar{n}^{n}\over(1+\bar{n})^{1+n}}\big|n\big\rangle\big\langle n\big|,\end{align} where $\bar{n}$ is the mean photon number.
Since the chaotic light does not exhibit correlation, for two-mode light, the density operator  can be treated as separable;
\begin{align}\label{s04}\hat{\rho}_{ab}=\sum_{n_{a},n_{b}=0}^{\infty}{\bar{n}_{a}^{n_{a}}\bar{n}_{b}^{n_{b}}
|n_{a},n_{b}\rangle\langle n_{a},n_{b}|\over(1+\bar{n}_{a})^{1+n_{a}}(1+\bar{n}_{b})^{1+n_{b}}},\end{align} 
where $\bar{n}_{a}$ and $\bar{n}_{b}$ are the mean photon numbers corresponding to the respective modes.

On the other hand, the number state can be describable in terms of the vacuum state as
\begin{align}\label{s05}|n\rangle={(\hat{a}^{\dagger})^{n}\over\sqrt{n}}\big|0\big\rangle,\end{align} which can also be expressed in terms of the coherent state
using the power series expansion of the form $e^{\alpha\hat{a}^{\dagger}}=\sum_{n=0}^{\infty}{(\alpha\hat{a}^{\dagger})^{n}\over n!}$ as
\begin{align}\label{s06}|\alpha\rangle=e^{-{\alpha^{*}\alpha\over2}}\sum_{n=0}^{\infty}{\alpha^{n}\over\sqrt{n!}}\left|n\right\rangle.\end{align}

Applying the density operator \eqref{s04} along with the relation of number state with coherent state \eqref{s06} and definition of the mean value  $\langle O\rangle=Tr(\hat{O}\hat{\rho})$, for any operator $\hat{O}$, it is possible to verify for two-mode chaotic light that
\begin{align}\label{s07}\langle\alpha^{*}\alpha\rangle=\bar{n}_{a},\end{align}
\begin{align}\label{s08}\langle\beta^{*}\beta\rangle=\bar{n}_{b},\end{align}
\begin{align}\label{s09}\langle\alpha\beta\rangle=\langle\alpha^{2}\rangle=\langle\beta^{2}\rangle=\langle\alpha\rangle=\langle\beta\rangle=0.\end{align}  

If it is found necessary, the mean photon number of each mode can be related to the frequency of the corresponding mode and temprature of the wall by 
\begin{align}\label{s10}\bar{n}_{i}={1\over e^{\omega_{i}/\kappa_{B} T}-1},\end{align} with $i=a,b$.

\subsection{Two-photon correlated emission laser}

A two-photon correlated emission laser can be interpreted as the amplification of simultaneously emitted photons with different frequency that are correlated due to the coherence superposition induced by preparing the atoms in the upper  and lower energy levels denoted by $|a\rangle$ and $|c\rangle$, respectively. In establishing the lasing mechanism, the atoms are presumed to be injected with constant rate $r_{a}$ into the cavity and left after successfully decayed to other energy levels. 

In general, interaction of a nondegenerate three-level cascade atom with a two-mode
cavity radiation can be described in the rotating-wave approximation and
the interaction picture by the Hamiltonian of the form
\begin{align}\label{s11}\hat{H}_{AR}=ig\big[\hat{a}|a\rangle\langle b|-|b\rangle\langle a|\hat{a}^{\dagger} -|c\rangle\langle b|\hat{b}^{\dagger} +\hat{b}|b\rangle\langle c|\big],\end{align} where $g$ is
the coupling constant taken to be the same for both transitions, 
$\hat{a}$ ($\hat{b}$) is the annihilation operator for one of the cavity radiations and $|b\rangle$ is the intermediate energy level.  

The atoms are initially prepared in arbitrary coherent superposition of energy levels between which a direct spontaneous transition is electric dipole forbidden, that is, the initial state of the
atom is chosen to be $|\Phi_{A}(0)\rangle\; = C_{a}(0)|a\rangle +
C_{c}(0)|c\rangle$, where $C_{a}(0)$ and $C_{c}(0)$ are probability amplitudes for the atoms to be initially in the upper and lower energy levels. The corresponding initial
density operator would be
\begin{align}\label{s12}\hat{\rho}_{A}(0) =\rho_{aa}^{(0)}|a\rangle\langle a| +\rho_{ac}^{(0)}|a\rangle\langle c| + \rho_{ca}^{(0)}|c\rangle\langle a| +\rho_{cc}^{(0)}|c\rangle\langle c|,\end{align} where
$\rho_{aa}^{(0)}$ and $\rho_{cc}^{(0)}$ are the corresponding populations and $\rho_{ac}^{(0)}$ represents the initial atomic coherence.

With this condition, applying the linear and
adiabatic approximation schemes  in the good cavity limit, the time
evolution of the reduced density operator that accounts for the
contribution of the cavity radiation due to the atoms is found to be \cite{pra79033810}
\begin{align}\label{s13}\frac{d\hat{\rho}}{dt} &= \frac{A(1-\eta)}{4} \big[2\hat{a}^{\dagger}\hat{\rho}\hat{a} -
\hat{a}\hat{a}^{\dagger}\hat{\rho} -\hat{\rho}\hat{a}\hat{a}^{\dagger}\big] \notag\\&+
\frac{A(1+\eta)}{4}\big[2\hat{b}\hat{\rho}\hat{b}^{\dagger} -\hat{b}^{\dagger}\hat{b}\hat{\rho} -
\hat{\rho}\hat{b}^{\dagger}\hat{b}\big] \notag\\& +\frac{A\sqrt{1-\eta^{2}}}{4}\big[\hat{\rho}\hat{a}^{\dagger}\hat{b}^{\dagger}
-2\hat{a}^{\dagger}\hat{\rho}\hat{b}^{\dagger} +\hat{a}^{\dagger}\hat{b}^{\dagger}\hat{\rho}\notag\\&-
2\hat{b}\hat{\rho}\hat{a} + \hat{a}\hat{b}\hat{\rho} +\hat{\rho}\hat{a}\hat{b}\big],\end{align} where $A={2r_{a}g^{2}\over\gamma^{2}}$ is the linear gain coefficient, $\eta=\rho_{cc}^{(0)}-\rho_{aa}^{(0)}$ is the population inversion and $\gamma$ is the atomic decay rate.

\section{Equations of evolution of the combined system}

In order to study the dynamics of the cavity radiation of the combined system, it is necessary to obtain the associated equations of evolution. To this end, the contribution of the initial chaotic light in the cavity and the two-mode vacuum reservoir towards the master equation is sought for. 

It is a well established fact that the time evolution of the reduced density operator for the cavity radiation
coupled to a reservoir has, in the Born approximation \cite{lou}, the form
\begin{align}\label{s14}\frac{d\hat{\rho}(t)}{dt} &= -i[\hat{H}_{S},\;\hat{\rho}(t)] -
i[\langle\hat{H}_{SR}(t)\rangle_{R},\; \hat{\rho}(0)] \notag\\&-
\int_{0}^{t}[\langle\hat{H}_{SR}(t)\rangle_{R},\;
[\hat{H}_{S}(t'), \;\hat{\rho}(t')]]dt' \notag\\&-
\int_{0}^{t}Tr_{R}[\hat{H}_{SR}(t),\; [\hat{H}_{SR}(t'),\;
\hat{\rho}(t')\hat{R}]]dt',\end{align} where $S$ and $R$ refer to
the system and reservoir variables and $\hat{\rho}(0)$ represents the radiation initially in the cavity. 

Assuming that the already heated cavity is placed in a vacuum (environment modes with sufficiently small photon number to disturb the analysis) and the injection of the atoms begins right after that, one can designate $\hat{\rho}(0)$ as a density operator for the chaotic light in the cavity.   Furthermore, the interaction of a two-mode cavity radiation with a two-mode reservoir can be
described in the interaction picture by the Hamiltonian
\begin{align}\label{s15}\hat{H}_{SR}(t) &=
i\sum_{k}\lambda_{k}[\hat{a}^{\dagger}\hat{a}_{k}e^{i(\omega_{0}-\omega_{k})t}
- \hat{a}\hat{a}^{\dagger}_{k}e^{-i(\omega_{0}-\omega_{k})t} \notag\\&+
\hat{b}^{\dagger}\hat{b}_{k}e^{i(\omega_{0}-\omega_{k})t} -
\hat{b}\hat{b}^{\dagger}_{k}e^{-i(\omega_{0}-\omega_{k})t}],\end{align}
where $\omega_{0}={\omega_{a}+\omega_{b}\over2}$, with
$\omega_{a}$ and $\omega_{b}$ stand for the frequencies of the cavity radiations, $(\hat{a}_{k},\hat{b}_{k})$ are the
annihilation operators, $w_{k}$ is the frequency and
$\lambda_{k}$ is the coupling constant for the $k^{th}$ mode
of the reservoir. 

In view of Eq. \eqref{s15}, one can write
\begin{align}\label{s16}\langle\hat{H}_{SR}(t)\rangle_{R}& =
i\sum_{k}\lambda_{k}[\hat{a}^{\dagger}
\langle\hat{a}_{k}\rangle_{R}e^{i(\omega_{0}-\omega_{k})t} \notag\\&-
\hat{a}\langle\hat{a}^{\dagger}_{k}\rangle_{R}e^{-i(\omega_{0}-\omega_{k})t}+
\hat{b}^{\dagger}\langle\hat{b}_{k}\rangle_{R}e^{i(\omega_{0}-\omega_{k})t}
\notag\\&-\hat{b}\langle\hat{b}^{\dagger}_{k}\rangle_{R}e^{-i(\omega_{0}-\omega_{k})t}].\end{align}
For a two-mode vacuum reservoir, 
\begin{align}\label{s17}\langle\hat{a}_{k}\rangle_{R} = \langle\hat{b}_{k}\rangle_{R} =0,\end{align} which leads to
\begin{align}\label{s18}\langle\hat{H}_{SR}\rangle_{R} = 0.\end{align}

This entails that the thermal light in the cavity does not directly contribute to the master equation. Hence solving the remaining terms following the standard approach \cite{lou} yields
\begin{align}\label{s19}\frac{d\hat{\rho}}{dt} &= \frac{\kappa}{2}
\big[2\hat{a}\hat{\rho}\hat{a}^{\dagger} -
\hat{a}^{\dagger}\hat{a}\hat{\rho} -
\hat{\rho}\hat{a}^{\dagger}\hat{a}\big] \notag\\&+
\frac{A(1-\eta)}{4} \big[2\hat{a}^{\dagger}\hat{\rho}\hat{a} -
\hat{a}\hat{a}^{\dagger}\hat{\rho} -
\hat{\rho}\hat{a}\hat{a}^{\dagger}\big] \notag\\&+
\frac{1}{4}\big[A(1+\eta)
+2\kappa\big]\big[2\hat{b}\hat{\rho}\hat{b}^{\dagger} -
\hat{b}^{\dagger}\hat{b}\hat{\rho} -
\hat{\rho}\hat{b}^{\dagger}\hat{b}\big] \notag\\ &+
\frac{A\sqrt{1-\eta^{2}}}{4}\big[\hat{\rho}\hat{a}^{\dagger}\hat{b}^{\dagger}
-2\hat{a}^{\dagger}\hat{\rho}\hat{b}^{\dagger} +
\hat{a}^{\dagger}\hat{b}^{\dagger}\hat{\rho}\notag\\&-
2\hat{b}\hat{\rho}\hat{a} + \hat{a}\hat{b}\hat{\rho} +
\hat{\rho}\hat{a}\hat{b}\big],\end{align} where $\kappa$ is the cavity damping constant.

Employing the master equation \eqref{s19} and following a straightforward algebra, it is possible to obtain
\cite{pra74043816,pra77013815}
\begin{align}\label{s20}\alpha(t)& = A_{+}(t)\alpha(0) +
B_{+}(t)\beta^{*}(0) + C_{+}(t)+D_{+}(t),\end{align}
\begin{align}\label{s21}\beta(t) &= A_{-}(t)\beta(0) +
B_{-}(t)\alpha^{*}(0) + C_{-}(t)+D_{-}(t),\end{align} where
 \begin{align}\label{s22}A_{\pm}(t) = \frac{1}{2\eta}\big[(\eta\mp1)e^{-\frac{\kappa+A\eta}{2}t} + (\eta\pm1)e^{-\frac{\kappa}{2}t}\big],\end{align}
 \begin{align}\label{s23}B_{\pm}(t) = \pm\frac{\sqrt{1-\eta^{2}}}{2\eta}\big[e^{-\frac{\kappa+A\eta}{2}t} -
 e^{-\frac{\kappa}{2}t}\big],\end{align}
\begin{align}\label{s24}C_{+}(t) &=
\frac{1}{2\eta}\int_{0}^{t}\big[(\eta-1)e^{-\frac{\kappa+A\eta}{2}(t-t')}
\notag\\&+ (\eta +1)e^{-\frac{\kappa}{2}(t-t')}\big]
f_{a}(t')dt',\end{align}
\begin{align}\label{s25}C_{-}(t) &=
\frac{1}{2\eta}\int_{0}^{t}\big[(\eta+1)e^{-\frac{\kappa+A\eta}{2}(t-t')}
\notag\\&+ (\eta -1)e^{-\frac{\kappa}{2}(t-t')}\big]
f_{b}(t')dt',\end{align}
\begin{align}\label{s26}D_{+}(t) &=
\frac{\sqrt{1-\eta^{2}}}{2\eta}\int_{0}^{t}\big[e^{-\frac{\kappa+A\eta}{2}(t-t')}
\notag\\&- e^{-\frac{\kappa}{2}(t-t')}\big]
f_{b}^{*}(t')dt',\end{align} 
\begin{align}\label{s27}D_{-}(t) &=
-\frac{\sqrt{1-\eta^{2}}}{2\eta}\int_{0}^{t}\big[e^{-\frac{\kappa+A\eta}{2}(t-t')}
\notag\\&- e^{-\frac{\kappa}{2}(t-t')}\big]
f_{a}^{*}(t')dt'\end{align} in which
\begin{align}\label{s28}\langle
f_{a}(t')f^{*}_{a}(t)\rangle =
{A(1-\eta)\over2}\delta(t-t'),\end{align}
\begin{align}\label{s29}\langle f_{b}(t')f_{a}(t)\rangle=
\frac{A\sqrt{1-\eta^{2}}}{4}\delta(t-t'),\end{align}
\begin{align}\label{s30}\langle f_{b}(t')f^{*}_{b}(t)\rangle &=\langle
f_{b}(t')f^{*}_{a}(t)\rangle =\langle f_{b}(t')f_{b}(t)\rangle
\notag\\&=\langle f_{a}(t')f_{a}(t)\rangle = 0.\end{align}

It is worth noting that  $\alpha(0)$ and $\beta(0)$ along with their corresponding complex conjugates represent the thermal light. It is clearly seen from Eqs. \eqref{s20} and \eqref{s21} that the thermal light enter into the evolution of the system as an initial condition as anticipated. But if the source of the chaotic light remains in the cavity throughout the lasing operation, the effect of the thermal light should also be included in the correlations of the noise forces. 

\section{Evolution of the two-mode squeezing and mean photon number}

\subsection{Two-mode squeezing}

Without lose of generality, a two-mode cavity radiation can be described by an
operator
\begin{align}\label{s31}\hat{c}={1\over\sqrt{2}}\big(\hat{a}+\hat{b}\big).\end{align}
Particularly, the squeezing properties of the
cavity radiation can be studied applying the quadrature operators
defined by
\begin{align}\label{s33}\hat{c}_{+}=\hat{c}^{\dagger}+\hat{c}\end{align}
and
\begin{align}\label{s34}\hat{c}_{-}=i(\hat{c}^{\dagger}-\hat{c}),\end{align} with the commutation relation 
\begin{align}\label{s35}\big[\hat{c}_{+},\;\hat{c}_{-}\big]=2i.\end{align}
Applying Eqs. \eqref{s31},  \eqref{s33}, \eqref{s34} and \eqref{s35}, the quadrature variances in terms of the $c$-number variables associated with the normal ordering are found to be
\begin{align}\label{s36}\Delta c_{\pm}^{2}& =1+ \frac{1}{2}\big[2\langle\alpha^{*}\alpha\rangle
\pm\langle\alpha^{*^{2}}\rangle \pm \langle\alpha^{2}\rangle   -
\langle\alpha^{*}\rangle^{2} - \langle\alpha\rangle^{2} \notag\\&\pm
\langle\beta^{*^{2}}\rangle \pm \langle\beta^{2}\rangle +
2\langle\beta^{*}\beta\rangle - \langle\beta^{*}\rangle^{2} -
\langle\beta\rangle^{2}\big] \pm \langle\alpha^{*}\beta^{*}\rangle\notag\\&
+ \langle\alpha^{*}\beta\rangle + \langle\alpha\beta^{*}\rangle
\pm \langle\alpha\beta\rangle \mp
\langle\alpha^{*}\rangle\langle\alpha\rangle-
\langle\alpha^{*}\rangle\langle\beta^{*}\rangle \notag\\&\mp
\langle\alpha^{*}\rangle\langle\beta\rangle \mp
\langle\alpha\rangle\langle\beta^{*}\rangle -
\langle\alpha\rangle\langle\beta\rangle \mp
\langle\beta^{*}\rangle\langle\beta\rangle. \end{align}

The various correlations in Eq. \eqref{s36} can be
evaluated applying Eqs. \eqref{s20} and \eqref{s21}. Thus on
the basis of Eqs. \eqref{s20} and  \eqref{s21}, and assuming the
cavity modes to be initially in a two-mode thermal light, one can
readily verify that
\begin{align}\label{s37}\langle\alpha(t)\rangle =
 \langle\beta(t)\rangle = 0.\end{align} 
Moreover, upon taking Eq. \eqref{s20} into account, 
\begin{align}\label{s38}\langle\alpha^{*}(t)\alpha(t)\rangle &=
A_{+}(t)\langle\alpha^{*}(t)\alpha(0)\rangle+B_{+}(t)\langle\alpha^{*}(t)\beta^{*}(0)\rangle
\notag\\&+ \langle\alpha^{*}(t)C_{+}(t)\rangle +
\langle\alpha^{*}(t)D_{+}(t)\rangle.\end{align} In line with this, using the complex
conjugate of Eq. \eqref{s20}, it is possible to see that
\begin{align}\label{s39}\langle\alpha^{*}(t)\alpha(0)\rangle &=
A_{+}(t)\langle\alpha^{*}(0)\alpha(0)\rangle+B_{+}(t)\langle\alpha(0)\beta(0)\rangle
\notag\\&+ \langle\alpha(0)C^{*}_{+}(t)\rangle +
\langle\alpha(0)D^{*}_{+}(t)\rangle.\end{align} 

Based on the fact that the noise force at time $t$ does not affect the
cavity mode variables at earlier times and taking the cavity modes to be initially in a chaotic  state, one gets
\begin{align}\label{s40}\langle\alpha^{*}(t)\alpha(0)\rangle
=A_{+}(t)\bar{n}_{a}.\end{align} It is also possible to verify in a similar manner
that
\begin{align}\label{s41}\langle\alpha^{*}(t)\beta^{*}(0)\rangle
=B_{+}(t)\bar{n}_{b},\end{align}
\begin{align}\label{s42}\langle\alpha^{*}(t)C_{+}(t)\rangle
=\langle C^{*}_{+}(t)C_{+}(t)\rangle+\langle
D^{*}_{+}(t)C_{+}(t)\rangle,\end{align}
\begin{align}\label{s43}\langle\alpha^{*}(t)D_{+}(t)\rangle
=\langle C^{*}_{+}(t)D_{+}(t)\rangle+\langle
D^{*}_{+}(t)D_{+}(t)\rangle.\end{align} 

Hence upon substituting
Eqs. \eqref{s40}, \eqref{s41}, \eqref{s42} and \eqref{s43} into
\eqref{s38}, one obtains
\begin{align}\label{s44}\langle\alpha^{*}(t)\alpha(t)\rangle &= \langle
C_{+}^{*}(t)C_{+}(t)\rangle + \langle D_{+}^{*}(t)D_{+}(t)\rangle
\notag\\&+\langle C^{*}_{+}(t)D_{+}(t)\rangle + \langle
D_{+}^{*}(t)C_{+}(t)\rangle\notag\\&+A_{+}^{2}(t)\bar{n}_{a}+B_{+}^{2}(t)\bar{n}_{b}. \end{align}

Next the remaining correlation functions in Eq. \eqref{s44} would be evaluated. To
this end, it is possible to write having Eq. \eqref{s24} that
\begin{align}\label{s45}&\langle C_{+}^{*}(t)C_{+}(t)\rangle =
\frac{1}{4\eta^{2}}\int_{0}^{t}\int_{0}^{t}[(\eta-1)e^{-(\kappa+A\eta)(t-t')/2}
 \notag\\&+ (\eta+1)e^{-\kappa(t-t')/2}][(\eta-1)e^{-(\kappa+A\eta)(t-t'')/2}
 \notag\\&+ (\eta+1)e^{-\kappa(t-t'')/2}]\langle f_{a}(t'')f_{a}^{*}(t')\rangle
 dt'dt'', \end{align} from which follows using  Eq. \eqref{s28} 
\begin{align}\label{s46}\langle C_{+}^{*}(t)C_{+}(t)\rangle &= {A(1-\eta)\over\eta^{2}}\left[\frac{(\eta-1)^{2}(1 -
e^{-(\kappa+A\eta)t})}{16(\kappa+A\eta)}
\right.\notag\\&\left. + \frac{(\eta+1)^{2}(1-e^{-\kappa t})}{16\kappa} \right.\notag\\&\left.
 + \frac{(\eta^{2}-1)(1 - e^{-(2\kappa+A\eta)t/2})}{4(2\kappa+A\eta)}\right].\end{align}
 It is not difficult to demonstrate in a similar manner that
\begin{align}\label{s47}\langle D_{+}^{*}(t)D_{+}(t)\rangle = 0,\end{align}
\begin{align}\label{s48}&\langle C_{+}^{*}(t)D_{+}(t)\rangle =
\langle D_{+}^{*}(t)C_{+}(t)\rangle \notag\\&={A(1-\eta^{2})\over \eta^{2}}\left[\frac{(\eta-1)(1 -
e^{-(\kappa+A\eta)t})}{32(\kappa+A\eta)}
\right.\notag\\&\left.-\frac{(\eta+1)(1-e^{-\kappa t})}{32\kappa}
+ \frac{(1 - e^{-(2\kappa+A\eta)t/2})}
{8(2\kappa+A\eta)}\right].\end{align} 

Therefore, in view of Eqs.
\eqref{s22}, \eqref{s23}, \eqref{s44}, \eqref{s46}, \eqref{s47} and \eqref{s48}, one can readily establishes 
\begin{align}\label{s49}\langle\alpha^{*}(t)\alpha(t)\rangle &=-{A(1-\eta)^{2}\over4\eta(\kappa+A\eta)}(1-e^{-(\kappa+A\eta)t})\notag\\&+{A(1-\eta^{2})\over2\eta(2\kappa+A\eta)}(1-e^{-(2\kappa+A\eta)t/2})
\notag\\&+{\bar{n}_{a}\over4\eta^{2}}\left[(\eta-1)^{2}e^{-(\kappa+A\eta)t}+(\eta+1)^{2}e^{-\kappa t}\right.\notag\\&\left.+2(\eta^{2}-1)e^{-(2\kappa+A\eta)t/2}\right]+{\bar{n}_{b}(1-\eta^{2})\over4\eta^{2}}\notag\\&\times\left[e^{-(\kappa+A\eta)t}+e^{-\kappa t}-2e^{-(2\kappa+A\eta)t/2}\right],\end{align}
\begin{align}\label{s50}\langle\beta^{*}(t)\beta(t)\rangle& = -{A(1-\eta^{2})\over4\eta(\kappa+A\eta)}(1-e^{-(\kappa+A\eta)t})\notag\\&+{A(1-\eta^{2})\over2\eta(2\kappa+A\eta)}(1-e^{-(2\kappa+A\eta)t/2})\notag\\&+{\bar{n}_{b}\over4\eta^{2}}\left[(\eta+1)^{2}e^{-(\kappa+A\eta)t}+(\eta-1)^{2}e^{-\kappa t}\right.\notag\\&\left.+2(\eta^{2}-1)e^{-(2\kappa+A\eta)t/2}\right]+{\bar{n}_{a}(1-\eta^{2})\over4\eta^{2}}\notag\\&\times\left[e^{-(\kappa+A\eta)t}+e^{-\kappa t}-2e^{-(2\kappa+A\eta)t/2}\right],\end{align}
\begin{align}\label{s51}\langle\alpha(t)\beta(t)\rangle &=
-{A(1-\eta)\sqrt{1-\eta^{2}}\over4\eta(\kappa+A\eta)}(1-e^{-(\kappa+A\eta)t})\notag\\&+{A\sqrt{1-\eta^{2}}\over2\eta(2\kappa+A\eta)}(1-e^{-(2\kappa+A\eta)t/2})\notag\\&-{\bar{n}_{a}\sqrt{1-\eta^{2}}\over4\eta^{2}}\left[(\eta-1)e^{-(\kappa+A\eta)t}\right.\notag\\&\left.-(\eta+1)e^{-\kappa t}+2e^{-(2\kappa+A\eta)t/2}\right]\notag\\&+{\bar{n}_{b}\sqrt{1-\eta^{2}}\over4\eta^{2}}\left[(\eta+1)e^{-(\kappa+A\eta)t}\right.\notag\\&\left.-(\eta-1)e^{-\kappa t}-2e^{-(2\kappa+A\eta)t/2}\right],\end{align}
\begin{align}\label{s52}\langle\alpha^{2}(t)\rangle=\langle\beta^{2}(t)\rangle=\langle\alpha^{*}(t)\beta(t)\rangle=0.\end{align}

On account of Eqs. \eqref{s37} and \eqref{s52}, the quadrature variances reduce to
\begin{align}\label{s53}\Delta c^{2}_{\pm}=1+\langle\alpha^{*}(t)\alpha(t)\rangle+\langle\beta^{*}(t)\beta(t)\rangle\pm2\langle\alpha(t)\beta(t)\rangle.\end{align}
In order to investigate the time development of the two-mode squeezing, applying Eqs. \eqref{s49}, \eqref{s50}, \eqref{s51} and \eqref{s52}, $\Delta c^{2}_{-}$ is plotted against $time$ for $\kappa=0.5$, $\eta=0.2$ and various values of $A$ given $\bar{n}_{a}$ and $\bar{n}_{b}$. The value of $\eta$ is fixed based on earlier study at steady state in which the degree of squeezing is found to be the highest at $\eta=0.2$ for $A=10$ \cite{pra74043816} and recent report for arbitrary time \cite{oc283781}.

\begin{figure}[hbt]
\centerline{\includegraphics [height=6.5cm,angle=0]{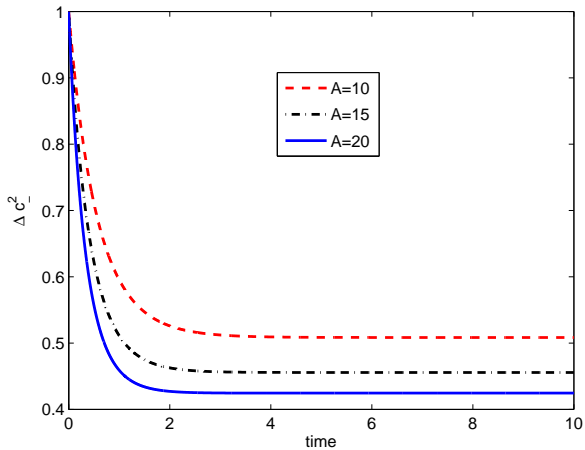}}
\caption {\label{fig1} Plots of the minus quadrature variance ($\Delta c_{-}^{2}$) of the cavity radiation for $\kappa=0.5$, $\bar{n}_{a}=\bar{n}_{b}=0$, $\eta=0.2$ and different values of $A$.} \end{figure}

\begin{figure}[hbt]
\centerline{\includegraphics [height=6.5cm,angle=0]{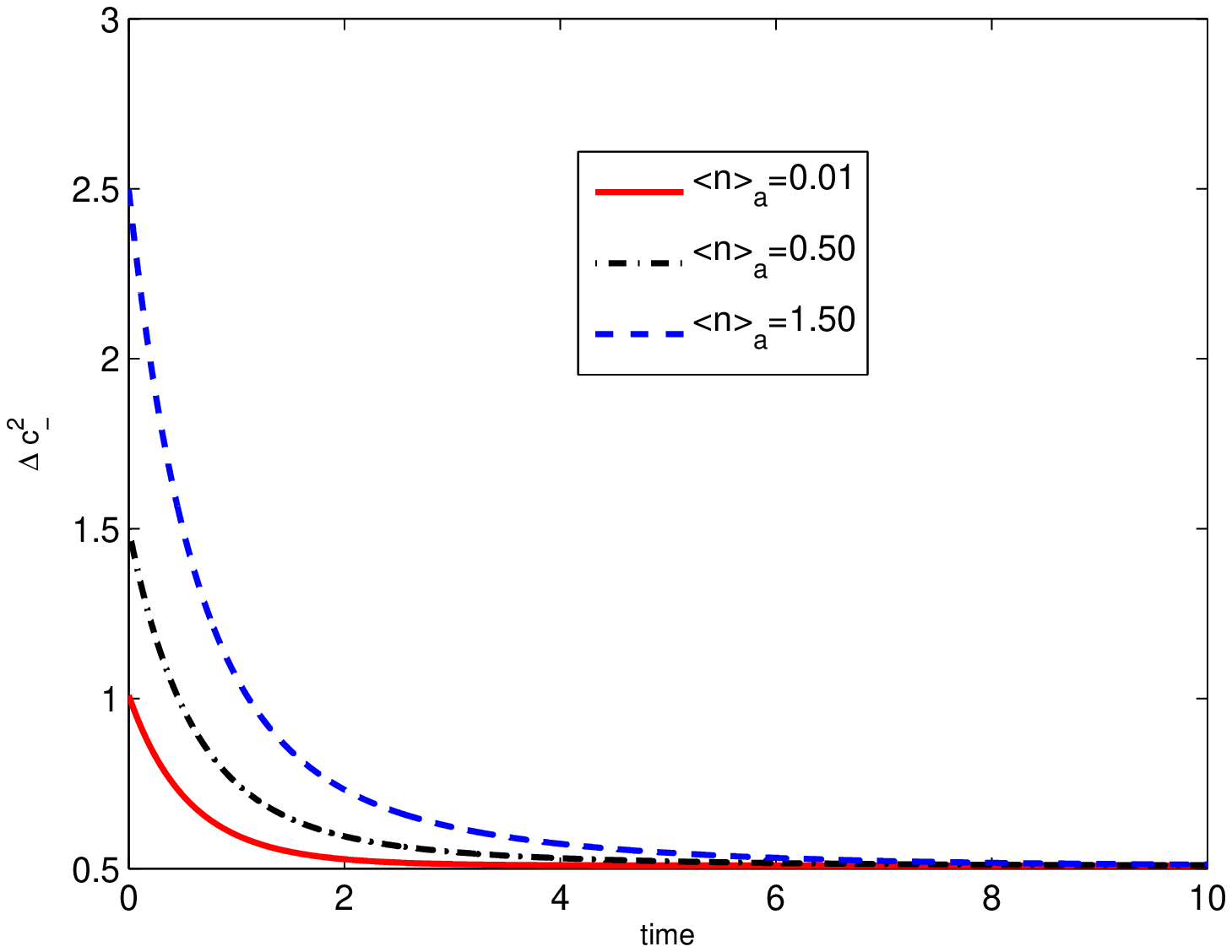}}
\caption {\label{fig2} Plots of the minus quadrature variance ($\Delta c_{-}^{2}$) of the cavity radiation for $\kappa=0.5$, $\bar{n}_{b}=0$, $\eta=0.2$, $A=10$ and different values of $\bar{n}_{a}$.} \end{figure}

\begin{figure}[hbt]
\centerline{\includegraphics [height=6.5cm,angle=0]{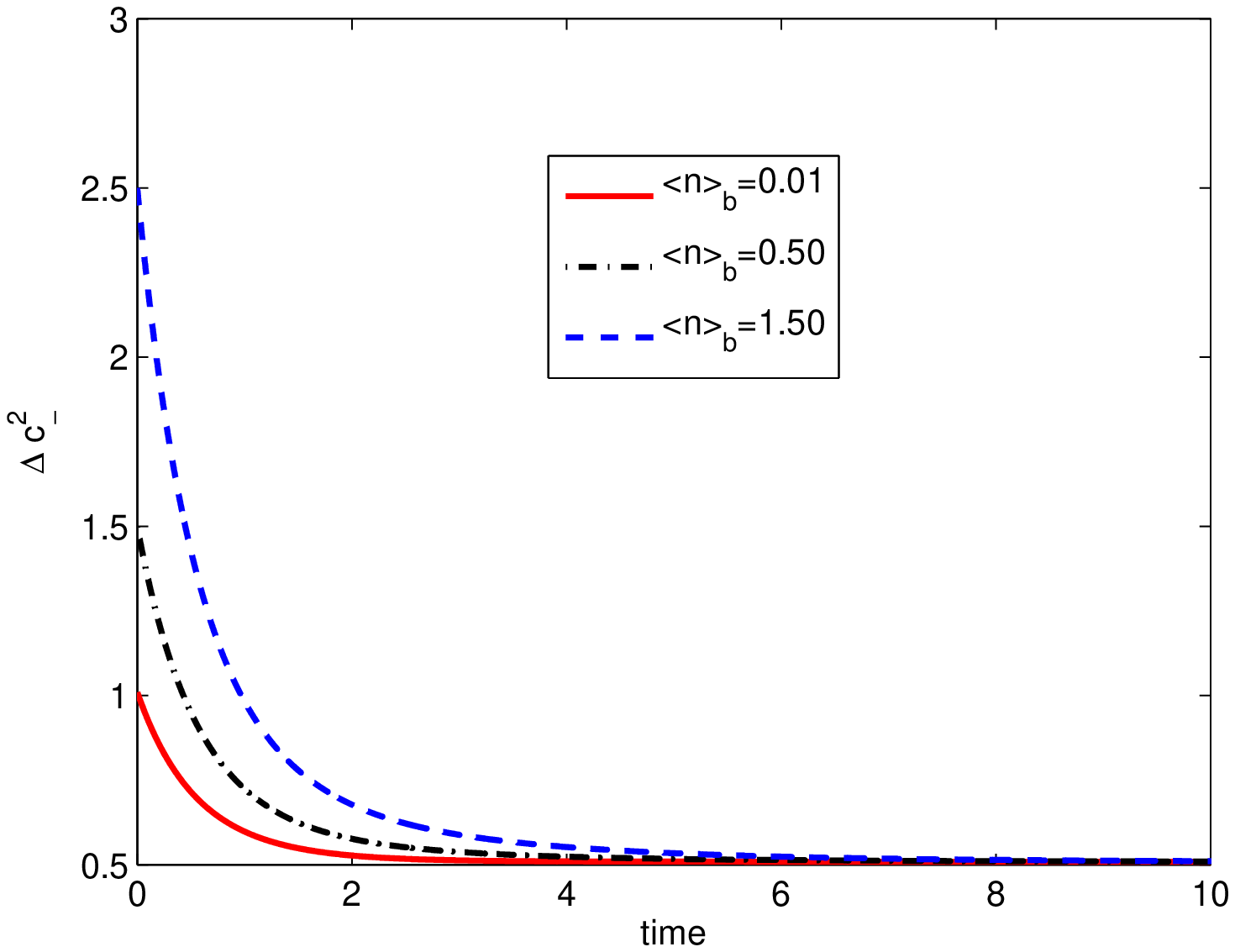}}
\caption {\label{fig3} Plots of the minus quadrature variance ($\Delta c_{-}^{2}$) of the cavity radiation for $\kappa=0.5$, $\bar{n}_{a}=0$, $\eta=0.2$, $A=10$ and different values of $\bar{n}_{b}$.} \end{figure}

\begin{figure}[hbt]
\centerline{\includegraphics [height=6.5cm,angle=0]{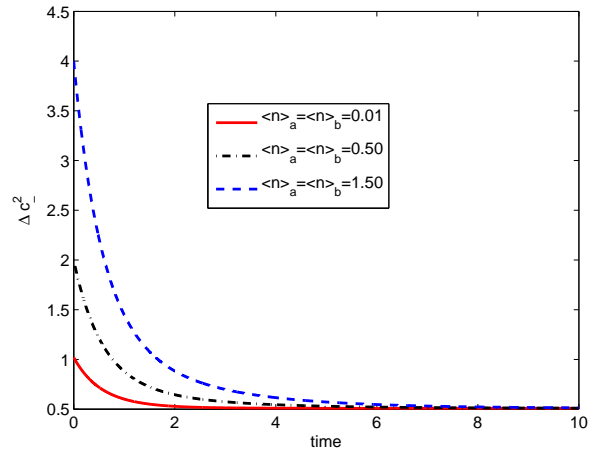}}
\caption {\label{fig4} Plots of the minus quadrature variance ($\Delta c_{-}^{2}$) of the cavity radiation for $\kappa=0.5$, $\eta=0.2$, $A=10$ and different values of $\bar{n}_{a}=\bar{n}_{b}$.} \end{figure}

It is clearly shown in Fig. \ref{fig1} that the degree of two-mode squeezing increases with the rate at which the atoms are injected into the cavity and time. This entails that with time more atoms get chance to traverse the cavity whereby the nonclassical correlation  acquires more strength. This type of reasoning may give sense if one compares the result obtained here with the earlier studies at steady state \cite{pra74043816}. It is also good to note that the degree of two-mode squeezing has peak value when $\eta=0$ and there is an external pumping radiation \cite{sint}. It has been argued elsewhere that such property of the two-mode squeezing is related to the thermal fluctuations resulting from vibrations of the atoms on the walls of the cavity. From reports at steady state, it is not difficult to observe that the mean photon number at $\eta=0$ is significantly larger than at $\eta=0.2$. In light of this, it is possible to envisage that the effect of the thermal fluctuations in the present case due to heating is not as big as when $\eta=0$ to overcome the correlation in extent that it reduces the degree of squeezing visibily. 

It goes beyond contention that the degree of squeezing in such a system is substantially limited by the thermal fluctuations specially at earlier stages and relaxed somewhat as time progresses since the strength of the correlation gets better as more atoms contribute towards establishing the correlation. Predominantely, it is the competition between the effect of the thermal fluctuations and the nonclassical correlation of the two modes that lead to the characteristic evolution of the two-mode squeezing and other nonclassical features. 

In order to understand the existing situation indepth, one needs to revisit the master equation \eqref{s19}. From the general form of the master equation, it is not difficult to infer that the second term stands for the gain of mode $a$, the third term for the lose of mode $b$ and the first term for the lose of mode $a$. It is also noteworthy that the first term and the part involving $\kappa$ in the third term comes from the contribution of the coupling of the cavity with external vacuum modes (reservoir) and basically related to the thermal fluctuations. On account of this, it may not be difficult to observe that until the lasing system able to generate strong light, the effect of the thermal fluctuations dominates and later, undoubtedly, the effect of the lasing system begins to overtake it. The result obtained here, the two-mode squeezing would be maximum at steady state, corroborates with this line of reasoning. 

It is vividly presented in Figs. \ref{fig2} and \ref{fig3} that the degree of two-mode squeezing is significantly impaired by chaotic light,  that is, the more intense the seeding, the more the squeezing is damaged. It turns out that there is no essential difference between when the initial thermal light is in mode $a$ and mode $b$. Referring to Eqs. \eqref{s49}, \eqref{s50} and \eqref{s51} bespeaks that at $t=0$, the thermal seed contributes to the variances of the quadrature operators ($\Delta c^{2}_{\pm}$) a term equal to $\bar{n}_{a}+\bar{n}_{b}$. In other words, at $t=0$, $\Delta c_{-}^{2}=1+\bar{n}_{a}+\bar{n}_{b}$. That is why, in the vicinity of $t=0$, the generated radiation would be in squeezed state only when the strength of the seed is significantly weaker. 

In the later stages of the process, the correlation of the emitted radiation would be strong enough to overcome the effect of the seed. If one critically observes Figs. \ref{fig2}, \ref{fig3} and \ref{fig4}, one may see that the degree of two-mode squeezing is almost the same for all cases after $t=2$. This suggests that the effect of the seed is relatively short lived.

Furthermore, making use of Eqs. \eqref{s49}, \eqref{s50}, \eqref{s51}  and \eqref{s53}, one readily finds at steady state
\begin{align}\label{s54}\Delta c_{\pm}^{2}&=
1+A(1-\eta)\notag\\&\times\left[\frac{4\kappa(\kappa+A\eta)+A[A+(2\kappa+A\eta)](1\pm\sqrt{1-\eta^{2}})}
{4\kappa(2\kappa+A\eta)(\kappa+A\eta)}\right]\notag\\&\pm
A\sqrt{1-\eta^{2}}\notag\\&\times\left[
\frac{4\kappa(\kappa+A\eta)+A^{2}(\eta^{2} - 1
\mp\sqrt{1-\eta^{2}})}
{4\kappa(2\kappa+A\eta)(\kappa+A\eta)}\right].\end{align}
It is not difficult to realize from Figs. \ref{fig2}, \ref{fig3}, \ref{fig4} and Eq. \eqref{s54} that the degree of squeezing at large time scale is independent of the mean photon number of the chaotic light. This is mainly because as time progresses there are more atoms that pass through the cavity in a way absorb the available radiation. Since the number of chaotic light in the cavity cannot increase, the absorption mechanism dwindles it over time. 

At this juncture, it would be appropriate noting that the atoms are continuously injected with constant rate and left the cavity after decaying to any other energy level that does not involve in the lasing process. In connection to this, what most probably happens is an atom after absorbing a chaotic light  emits a light which is not necessarily has the same nature as it absorbs. Since this process successively repeated, the chaotic light in the cavity is converted to a correlated light or lost all the way which leads to diminishing effect of the seed on the properties of the cavity radiation. 

\subsection{Mean photon number}

The mean number of photon pairs  describing the two-mode cavity
radiation can be defined as
\begin{align}\label{s55}\bar{N}=\langle\hat{c}^{\dagger}(t)\hat{c}(t)\rangle,\end{align}
where $\hat{c}(t)$ is the annihilation operator \eqref{s31}. It is readily noticeable that the operators in Eq. \eqref{s55} are in the normal order. Hence it is possible to express Eq. \eqref{s55} in terms
of $c$-number variables associated with the normal ordering as
\begin{align}\label{s56}\bar{N}={1\over2}\big[\langle\alpha^{*}(t)\alpha(t)\rangle+\langle\beta^{*}(t)\beta(t)\rangle\big],\end{align} where $2\bar{N}$ represents the mean photon number \cite{jpb41145501}. 

In the following, the mean photon number is plotted against $time$ for $\eta\approx0$, $\kappa=0.5$ and different values of  $A$, $\bar{n}_{a}$ and $\bar{n}_{b}$. Here the value of $\eta$ is chosen on the basis of earlier studies at steady state in which the mean photon number is found to be maximum for $\eta=0$. The value of $A$ on the other hand is fixed to be small so that the dynamics of the intensity of the radiation be evident from the plots. 

\begin{figure}[hbt]
\centerline{\includegraphics [height=6.5cm,angle=0]{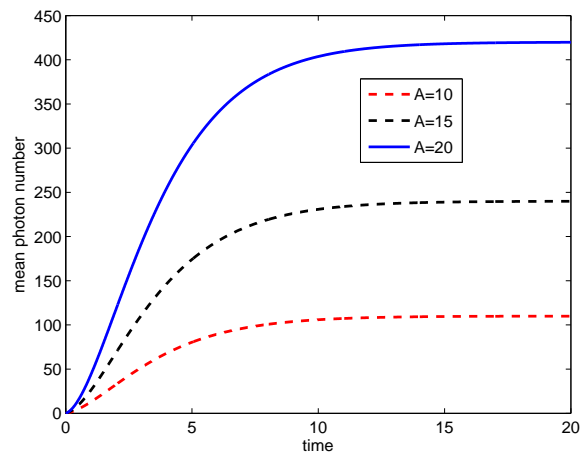}}
\caption {\label{fig5} Plots of the mean photon number of the cavity radiation ($2\bar{N}$) for $\kappa=0.5$, $\bar{n}_{a}=\bar{n}_{b}=0$, $\eta\approx0$, and different values of $A$.} \end{figure}

\begin{figure}[hbt]
\centerline{\includegraphics [height=6.5cm,angle=0]{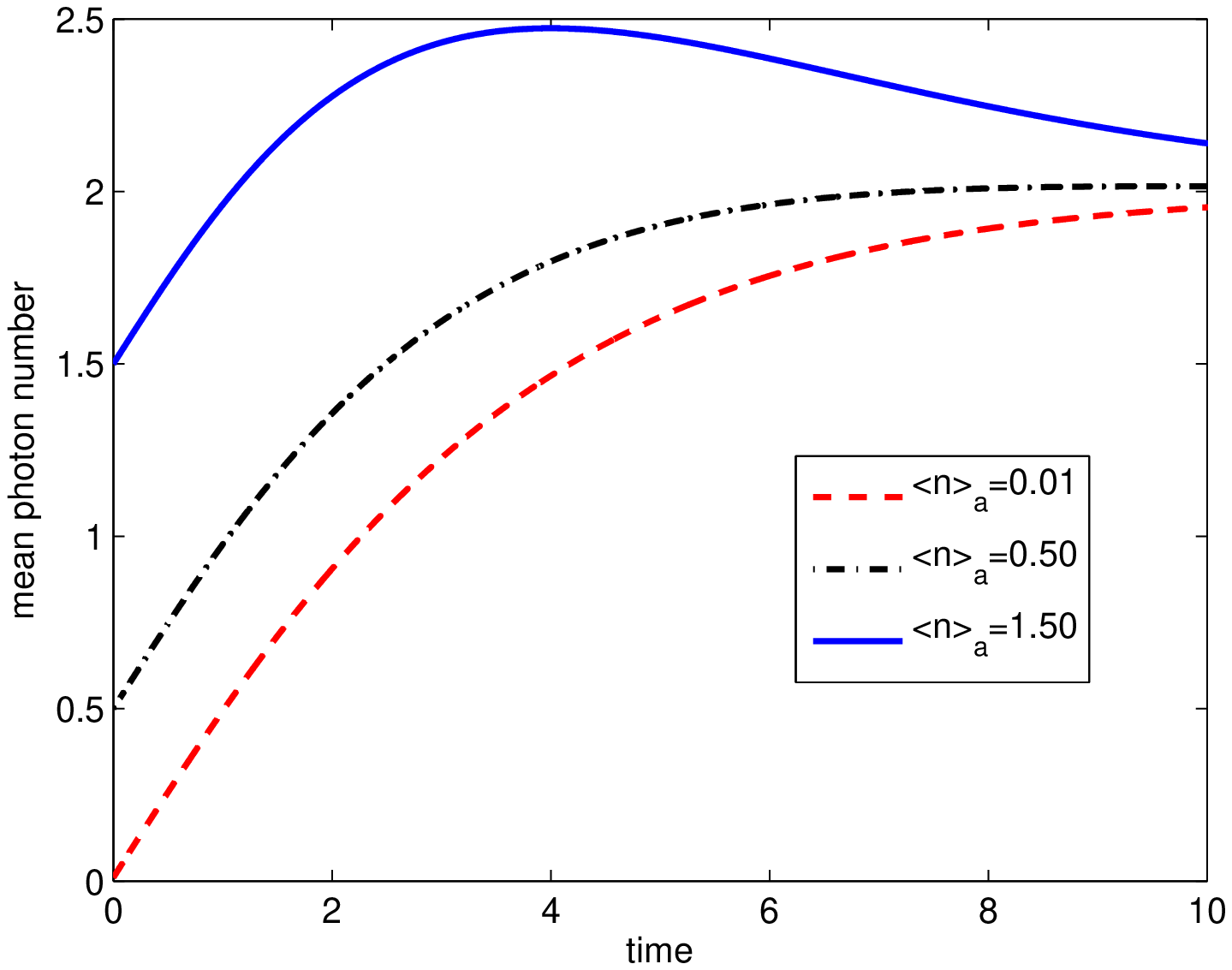}}
\caption {\label{fig6} Plots of the mean photon number of the cavity radiation ($2\bar{N}$)  for $\kappa=0.5$, $\bar{n}_{b}=0$, $\eta\approx0$, $A=1$, and different values of $\bar{n}_{a}$.} \end{figure}

\begin{figure}[hbt]
\centerline{\includegraphics [height=6.5cm,angle=0]{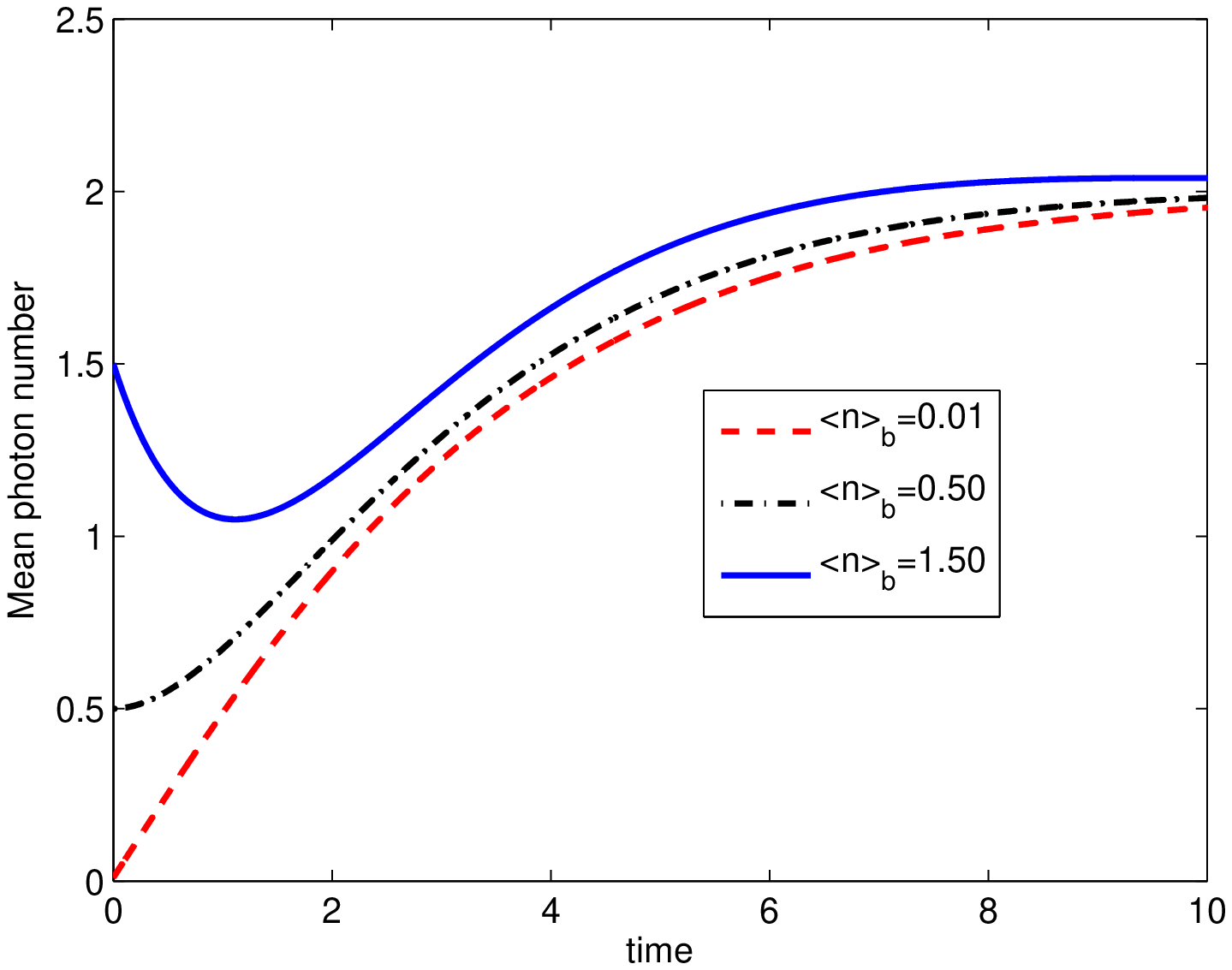}}
\caption {\label{fig7} Plots of the mean photon number of the cavity radiation ($2\bar{N}$)  for $\kappa=0.5$, $\bar{n}_{a}=0$, $\eta\approx0$, $A=1$, and different values of $\bar{n}_{b}$.} \end{figure}

\begin{figure}[hbt]
\centerline{\includegraphics [height=6.5cm,angle=0]{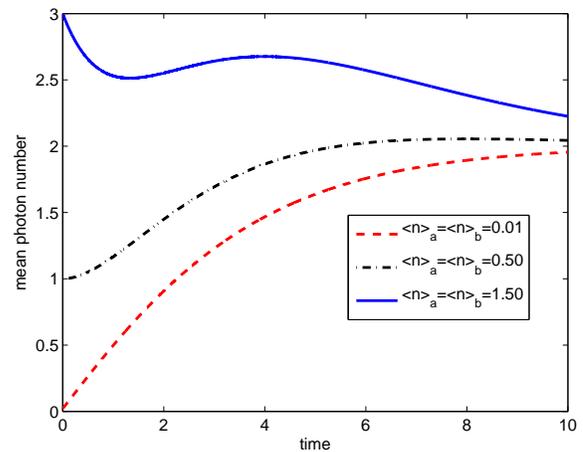}}
\caption {\label{fig8} Plots of the mean photon number of the cavity radiation ($2\bar{N}$)  for $\kappa=0.5$,  $\eta\approx0$, $A=1$, and different values of $\bar{n}_{a}=\bar{n}_{b}$.} \end{figure}

It is clearly shown in Fig. \ref{fig5} that the mean photon number increases with time and linear gain coefficient when there is no seeding. This is readily understandable since  traversing more atoms across the cavity most likely leads to generation of more photons. However, the results indicated in Figs. \ref{fig6}, \ref{fig7} and \ref{fig8} do not support this idea when there is a seed in general terms. 

To begin with, when $\bar{n}_{b}=0$ and $\bar{n}_{a}\ne0$, the situation indicated in Fig. \ref{fig6}, the mean photon number increases with time at the beginning and then decreases later. This is mainly related to the fact that at initial time the atoms emit until the radiation in the cavity is sufficient enough to frustrate the transition from energy level $|a\rangle$ to $|b\rangle$. Later due to the availability of large number of photons than the system can generate, the atoms begin to pickup these photons leisurely and underwent emissions to any other level that does not involve in the lasing mechanism. In the process, as earlier results display, the number of chaotic light starts to dwindle which compels the decline of the mean photon number. Critical scrutiny reveals that the mean photon number does not slump to the saturation value by more than the mean photon number initially in the cavity and  the mean photon number cannot be less than its value at $t=0$ for parameters under consideration.

Quite contrary to this, when $\bar{n}_{a}=0$ and $\bar{n}_{b}\ne0$, the situation dipcited in Fig. \ref{fig7}, the mean photon number initially decreases  and later starts to increase with time. In this case, it is worth noting that nearly 50\% of the atoms are prepared to be in the lower energy level and most probably hungry for photons.  As a result, the photons due to the seed which are available in the cavity readily absorbed by this atoms and consequently the number of the photon drastically falls. As critical scrutiny might have revealed, the depletion in mean photon number cannot go lower than the initial mean photon number for all conceivable cases. 

After quite sometime, the available photons ($\bar{n}_{b}$) begins to dwindle whereby the usual emission process gets a chance to regain control in which the mean photon number readily increases till it becomes nearly uniform at longer time scale. It is not difficult to observe from Fig. \ref{fig8} that specially when $\bar{n}_{a}=\bar{n}_{b}=1.5$, both these effects are readily observed. That means the mean photon number decreases with time, then increases and later decreases with time. This indicates that the absorption of mode $b$ is efficient in earlier times whereas that of mode $a$ is at later times. Almost similar behavior of mean photon number has been predicted for different parameters when the initial radiation is assumed to be a coherent light \cite{oc283781}.

\section{Quantification of entanglement}

Based on different physical contexts and mathematical considerations, several sufficient inseparability criteria for composite state have been proposed \cite{prl93063601,prl842722,prl95120502,pla2231,prl771413,prl842726,prl88120401,pra67052104,pra67022320,prl96050503,prl95230502} for continuous variables. One of these alternatives is the criterion associated with the logarithmic negativity \cite{pra70022318,pra65032314,job7577} that can be defined as
\begin{align}\label{s59}E_{N}=max[0,-\log_{2}V_{s}],\end{align} where $V_{s}$ is the smallest eigenvalue \cite{prl93063601,pra70022318}. 

Upon solving the eigenvalue Eqs. for symplectic spectrum of the covariance matrix of the partially transposed density operator, the smallest eigenvalue is found to have the form
\begin{align}\label{s60}V_{s}=\left[\frac{\sigma-\sqrt{\sigma^{2}-4det\Omega}}{2}\right]^{1/2},\end{align} with
$\sigma= det A+det B-2det C_{AB}$,
in which $A$ and $B$ are the covariance matrices that represent each modes separately while $C_{AB}$ is the intermodal correlations and $\Omega$ is the covariance matrix to be defined later \cite{prl93063601,pra65032314}. 

In line with this consideration, the $2X2$ block form of the covariance matrix  can be expressed  as
\begin{align}\label{s61}\Omega=\left(
                         \begin{array}{cc}
                           A & C_{AB} \\
                           C_{AB}^{T} & B \\
                         \end{array}
                       \right),
\end{align} where $\Omega_{ij}={1\over2}\langle\hat{X}_{i}\hat{X}_{j}+\hat{X}_{j}\hat{X}_{i}\rangle -\langle\hat{X}_{i}\rangle\langle\hat{X}_{j}\rangle$. It is essential to note that the involved quadrature operators can be defined as
$\hat{X}_{1}=\hat{a}+\hat{a}^{\dagger}$, $\hat{X}_{2}=i\big[\hat{a}^{\dagger}-\hat{a}\big]$,
$\hat{X}_{3}=\hat{b}+\hat{b}^{\dagger}$ and
$\hat{X}_{4}=i\big[\hat{b}^{\dagger}-\hat{b}\big]$. It is also worth noting that $\hat{X}_{i}$'s can be related to the quadrature operators by $\hat{c}_{+}={1\over\sqrt{2}}(\hat{X}_{1}+\hat{X}_{3})$ and $\hat{c}_{-}={1\over\sqrt{2}}(\hat{X}_{2}+\hat{X}_{4})$. 

With this introduction, the extended covariance matrix turns out to have the form
\begin{align}\label{s62}\Omega=\left(
                         \begin{array}{cccc}
                           m & 0 & c & 0 \\
                           0 & m & 0 & -c \\
                           c & 0 & n & 0 \\
                           0 & -c & 0 & n \\
                         \end{array}
                       \right),
\end{align} where $m=2\langle\hat{a}^{\dagger}(t)\hat{a}(t)\rangle +1$, $n=2\langle\hat{b}^{\dagger}(t)\hat{b}(t)\rangle +1$ and $c=\langle\hat{a}(t)\hat{b}(t)\rangle +\langle\hat{a}^{\dagger}(t)\hat{b}^{\dagger}(t)\rangle$ which can also be expressed in terms of $c$-number variables associated with the normal ordering as
$m=2\langle\alpha^{*}(t)\alpha(t)\rangle +1$,
$n=2\langle\beta^{*}(t)\beta(t)\rangle +1$ and
$c=\langle\alpha(t)\beta(t)\rangle +\langle\alpha^{*}(t)\beta^{*}(t)\rangle$.

The two-mode Gaussian composite state is entangled when the logarithmic negativity is positive, that is, if $E_{N}=-Log_{2}V_{s}$ which implies that $Log_{2}V_{s}$ should be negative. This, on the other hand, holds true provided \cite{prl93063601} that
\begin{align}\label{s63}V_{s}<1.\end{align}

Furthermore, making use  of Eqs. \eqref{s61}, \eqref{s62}, and the accompanying definitions of $m$, $n$ and $c$, one can readily see that
\begin{align}\label{s64}det A&=4\langle\alpha^{*}(t)\alpha(t)\rangle[\langle\alpha^{*}(t)\alpha(t)\rangle+1]+1,\end{align}
\begin{align}\label{s65}det B&=4\langle\beta^{*}(t)\beta(t)\rangle[\langle\beta^{*}(t)\beta(t)\rangle+1]+1,\end{align}
\begin{align}\label{s66}det C_{AB}&=-4\langle\alpha(t)\beta(t)\rangle^{2},\end{align}
\begin{align}\label{s67}det \Omega &= 16 \left[{1\over4}+{\langle\alpha^{*}(t)\alpha(t)\rangle+\langle\beta^{*}(t)\beta(t)\rangle\over2}
\right.\notag\\&\left.+\langle\alpha^{*}(t)\alpha(t)\rangle\langle\beta^{*}(t)\beta(t)\rangle-\langle\alpha(t)\beta(t)\rangle^{2}\right]^{2}.\end{align}

In order to study the evolution of the entanglement using criterion \eqref{s63}, the smallest eigenvalue ($V_{s}$) is plotted against $time$ applying Eqs. \eqref{s49},  \eqref{s50}, \eqref{s51}, \eqref{s63}, \eqref{s64}, \eqref{s65}, \eqref{s66}  and \eqref{s67}. Taking the value of $\eta$ to be very close to zero and similar values for the rest of the parameters as for two-mode squeezing, the evolution of the entanglement would be studied.

\begin{figure}[hbt]
\centerline{\includegraphics [height=6.5cm,angle=0]{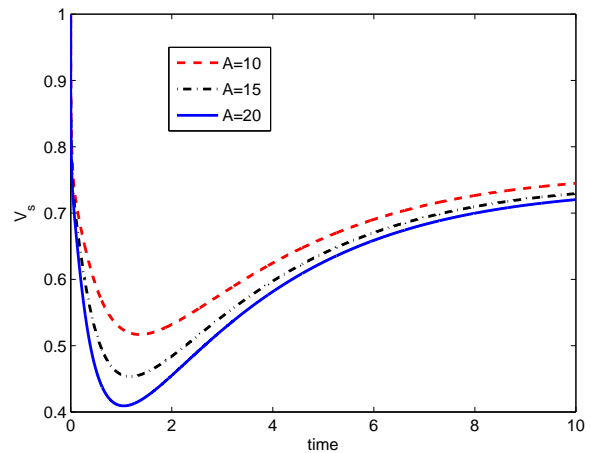}}
\caption {\label{fig9} Plots of the minimum of the logarithmic negativity ($V_{s}$) for $\kappa=0.5$,  $\eta\approx0$, $\bar{n}_{a}=\bar{n}_{b}=0$, and different values of $A$.} \end{figure}

\begin{figure}[hbt]
\centerline{\includegraphics [height=6.5cm,angle=0]{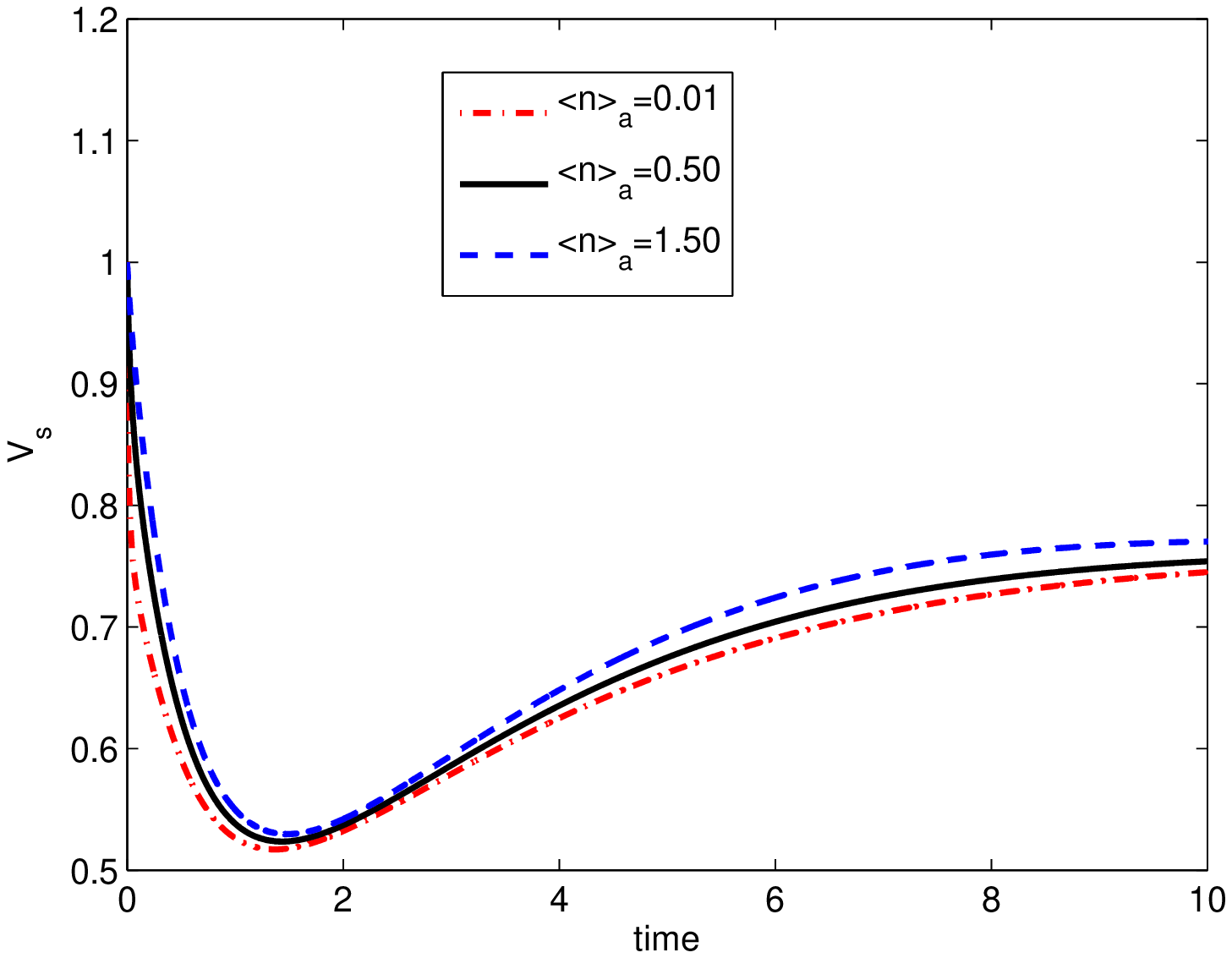}}
\caption {\label{fig10} Plots of the minimum of the logarithmic negativity ($V_{s}$) for $\kappa=0.5$,  $\eta\approx0$, $\bar{n}_{b}=0$, $A=10$,  and different values of $\bar{n}_{a}$.} \end{figure}

\begin{figure}[hbt]
\centerline{\includegraphics [height=6.5cm,angle=0]{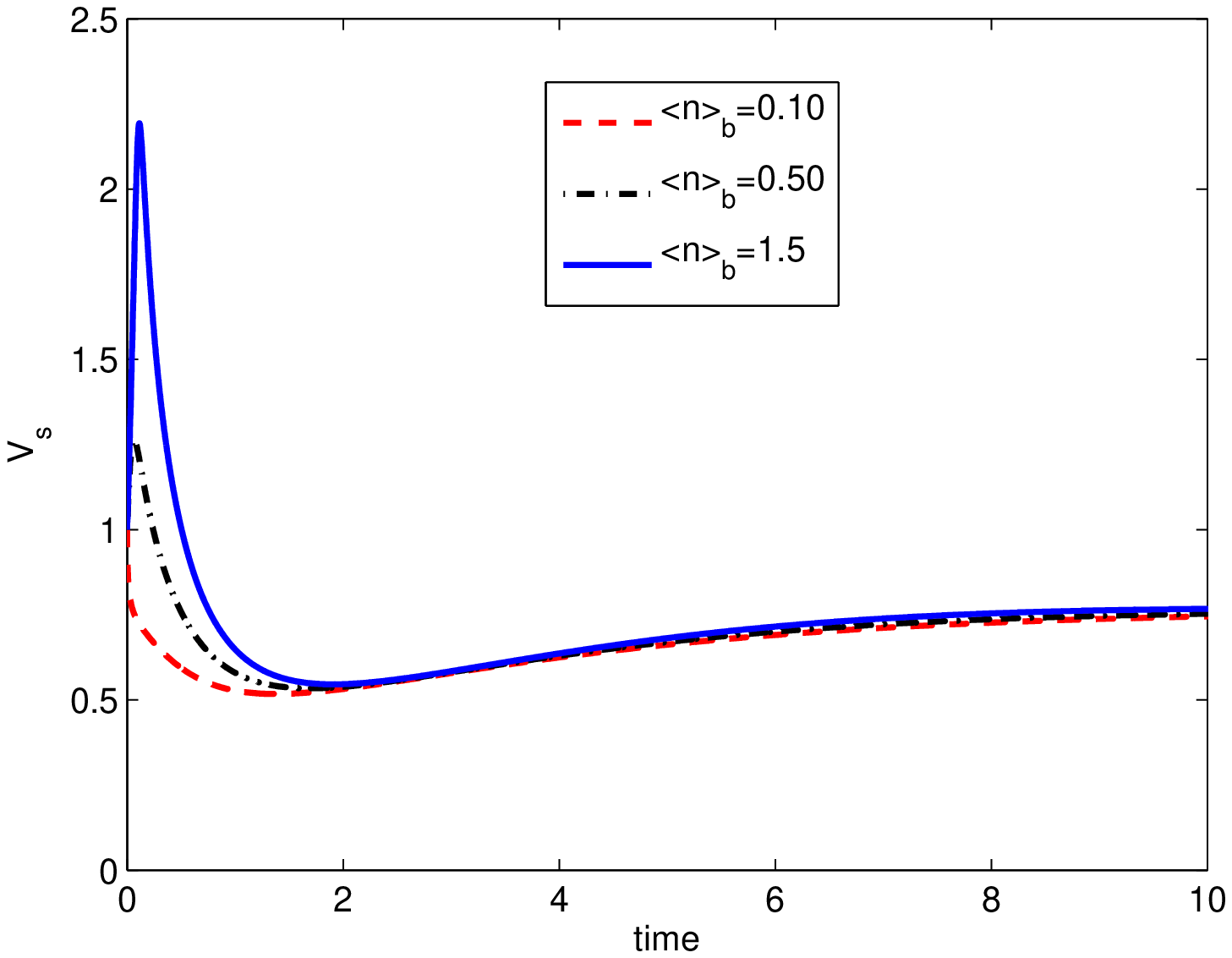}}
\caption {\label{fig11} Plots of the minimum of the logarithmic negativity ($V_{s}$) for $\kappa=0.5$,  $\eta\approx0$, $\bar{n}_{a}=0$, $A=10$, and different values of $\bar{n}_{b}$.} \end{figure}

\begin{figure}[hbt]
\centerline{\includegraphics [height=6.5cm,angle=0]{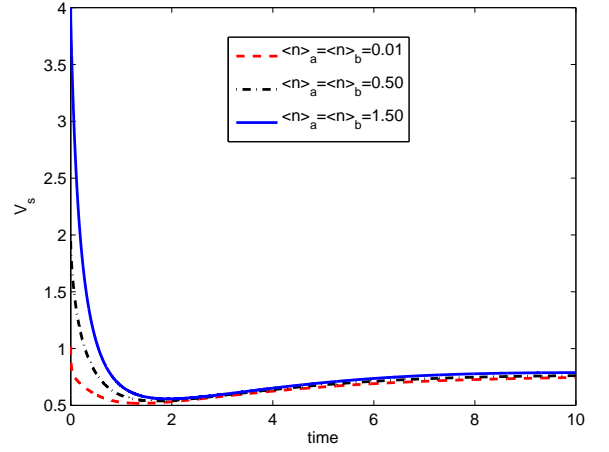}}
\caption {\label{fig12} Plots of the minimum of the logarithmic negativity ($V_{s}$) for $\kappa=0.5$,  $\eta\approx0$, $A=10$, and different values of $\bar{n}_{a}=\bar{n}_{b}$.} \end{figure}

It is clearly indicated in Fig. \ref{fig9} that the degree of entanglement increases with the linear gain coefficient for values under consideration. It so happens that the degree at which the criterion \eqref{s63} is violated increases with time at early stages of the operation, but it decreases in the later stages. Similar behavior of the nonclassical features including two-mode squeezing \cite{sint} and entanglement measure following from Duan {\it{et al.}} \cite{pra79013831,prl842722} when there is an external driving radiation and phase fluctuations was reported. 

Most recently, a thorough explanation in this regard is provided in relation to the thermal fluctuations due to vibrations of the atoms on the walls of the cavity. Complementary explanation is also provided in Section IV. Though it is not clearly observed here, earlier studies show that the minimum of the logarithmic negativity can go as small as 0.25 for $A=10$ at steady state \cite{jpb42215506}. Moreover, comparison of Figs. \ref{fig5} and \ref{fig9} reveals that $V_{s}$ attains minimum value (better degree of entanglement) quite earlier than when the intensity becomes strong. Therefore, apart from what is expected at steady state, as it stands, the practical utility of this system for generation of a strong entangled light does not seem satisfactory. 

Furthermore, it is not difficult to see from Figs. \ref{fig10} and \ref{fig11} that at the vicinity of $t=0$, the entanglement of the cavity radiation when $\bar{n}_{a}=0;~~\bar{n}_{b}\ne0$ and $\bar{n}_{a}\ne0;~~\bar{n}_{b}=0$ exhibits different nature. By large, this result can be associated with the nature of the mean photon number in a similar case. Even though the presence of thermal seed in mode $a$ somewhat diminishes the degree of entanglement, its effect would be more prominent when the thermal light is seeded in mode $b$. Critical comparison between Figs. \ref{fig9} and \ref{fig10}, for $A=10$, indicates that the effect of the thermal light when it is in mode $a$ is quite nominal over time, but as clearly shown in Fig. \ref{fig12} the effect would be large when $\bar{n}_{a}=\bar{n}_{b}$ is large specially at the beginning of the lasing process. In consonant with earlier observations, the effect of the thermal seeding would be insignificant in the later time of the operation to change the achievable entanglement. 

\begin{figure}[hbt]
\centerline{\includegraphics [height=6.5cm,angle=0]{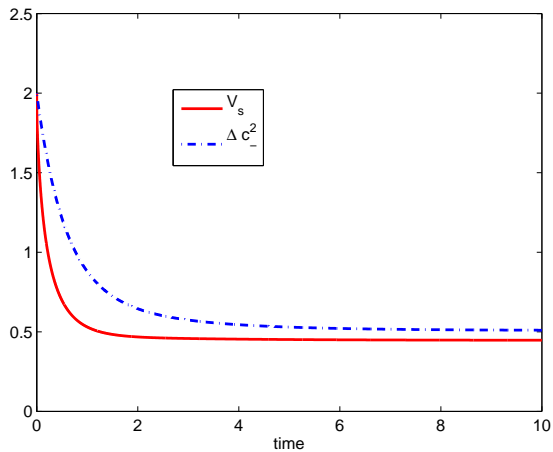}}
\caption {\label{fig13} Plots of the minimum of the logarithmic negativity ($V_{s}$) and the minus quadrature variance ($\Delta c_{-}^{2}$) for $\kappa=0.5$,  $\eta=0.2$, $A=10$, and $\bar{n}_{a}=\bar{n}_{b}$=0.5.} \end{figure}

In order to see the relation between two-mode squeezing and entanglement, $V_{s}$ and $\Delta c_{-}^{2}$ are plotted against $time$ for selected parameters. It is evident from Fig. \ref{fig13} that the nature of the evolution of the degree of entanglement and two-mode squeezing is basically similar, although observing entangled light when there is no squeezing at early stages of the process is possible. The resemblance as in Fig. \ref{fig13} remains appearing at larger time scale as predicted earlier at steady state \cite{jpb42215506}. It is worth noting that such a tendency of exhibiting similar behavior would break near $\eta=0$, since the squeezing disappears at $\eta=0$ while the degree of entanglement is nearly close to 50\%  which is undoubtedly a significant difference. It is due to this disparity that the values of $\eta$  are taken to be different while investigating two-mode squeezing  and entanglement.

\section{Conclusion}

In this contribution, detailed analysis of the time evolution of the cavity radiation of the two-photon correlated emission laser when the cavity is initially maintained at a particular temprature is presented. The effect of the seed thermal light is included via the parameters that describe the initial conditions in the solution of the corresponding stochastic differential equations. With the understanding that the effect of the thermal light enters as initial conditions only if the source is entirely shut off prior to the injection of the atoms,  the characteristic correlations of the chaotic light are employed in the process of determining various quantities of interest. 

It turns out that the thermal light significantly damages the nonclassical features like two-mode squeezing and entanglement at earlier stages of the lasing process. Although the degree of entanglement and two-mode squeezing  exhibit more or less a similar dependence on the intensity of the thermal light and time, their nature comes out to be different when the thermal light is seeded only in one of the modes specially at the vicinity of $t=0$. Fundamentally, the degree of two-mode squeezing is almost independent of which mode is initially seeded, but the degree of entanglement reduces considerably when the light with the same strength is seeded in mode $b$. 

Quite remarkably, a somewhat similar characteristic is perceived for the mean photon number. When viewed over a longer time span, the mean photon number has a peak when mode $a$ is seeded, but a dip when mode $b$ is seeded. The time for which one observes the peak is relatively much greater than the corresponding time for the dip. It is found that this disparity in the tendency of the mean photon number increases with the intensity of the seed.

Moreover, it is unambiguousely shown that at longer time scale, the two-mode squeezing, entanglement and intensity of the radiation are independent of the strength of the initial thermal light. This by large is related to  the presumption that the atoms can absorb any light in the cavity as long as it is resonant. In view of this, it would be an obvious matter to realize that the effect of the thermal light is sucked from the cavity in a level that it would not affect the nonclassical features of the radiation. 

As long as the radiation in the cavity is resonant with the radiative levels of the atoms, it is not difficult to infer that this outcome remains valid irrespective of the type of the seed. This  entails that while designing the experimental setup, one should not be compelled by the remnant radiation in the cavity, since its effect would be significantly nominal at longer time scales. It is hoped that such an outcome can provide another dimension to the practical utilization of the various schemes of the three-level laser as a source of bright entangled light.

\section*{Acknowledgment}

I thank the Max Planck Institute for the Physics of  Complex Systems for allowing me to visit them and use their facility in carrying out this research and Dilla University for granting the leave of absence. I am also grateful to my colleague Tewdros for his serious comments on the manuscript.


\end{document}